\documentclass[aps,prd,10pt,twocolumn,superscriptaddress,nofootinbib
  ,noshowpacs,preprintnumbers]{revtex4-1}
\usepackage{amsmath,amssymb}
\usepackage[version=4]{mhchem}
\usepackage{graphicx}
\usepackage{url}
\usepackage{color}
\usepackage{lipsum}
\usepackage[dvipsnames]{xcolor}
\usepackage[colorlinks=true,breaklinks=true]{hyperref}
\hypersetup{allcolors=[rgb]{0.0 0.0 0.6},linkcolor=[rgb]{0.75 0.05 0.05}}

\usepackage{multirow}

\newcommand{\exclude}[1]{}

\def\Z2{$\mathcal{Z_2}$}

\newcommand {\ignore}[1]{}

\begin{document}

%\author{Mario Reig}\email{mario.reig@ific.uv.es}
%\affiliation{\AddrAHEP}

\preprint{NORDITA-2020-033}

\title{Revisiting longitudinal plasmon-axion conversion in external magnetic fields}

%%%%%%%%%%%%%%%%%%%%%%%%%%%%%%%%%%%%%%%%%%%%%%%%%%%%%%%%%

%\begin{textblock*}{5cm}(11cm,-8.2cm)
%\end{textblock*}

\pacs{98.80.Cq, 14.80.Va, 12.10.Dm}
%%%%%%%%%%%%%%%%%%%%%%%%%%%%%%%%%%%%%%%%%%%%%%%%%%
\author{Andrea Caputo}\email{andrea.caputo@uv.es}\affiliation{Instituto de Fisica Corpuscular, Universidad de Valencia and CSIC, Edificio Institutos Investigacion, Catedratico Jose Beltran 2, Paterna, 46980 Spain}
\author{Alexander J. Millar}\email{alexander.millar@fysik.su.se}\affiliation{The Oskar Klein Centre for Cosmoparticle Physics,
Department of Physics,
Stockholm University, AlbaNova, 10691 Stockholm, Sweden}
\affiliation{Nordita, KTH Royal Institute of Technology and
Stockholm
  University, Roslagstullsbacken 23, 10691 Stockholm, Sweden}
\author{Edoardo Vitagliano}\email{edoardo@physics.ucla.edu}\affiliation{ Department  of  Physics  and  Astronomy,  University  of  California,  Los  Angeles,  California,  90095-1547,  USA}

\begin{abstract} 
In the presence of an external magnetic field the axion and the photon mix. In particular, the dispersion relation of a longitudinal plasmon always crosses the dispersion relation of the axion (for small axion masses), thus leading to a resonant conversion. Using thermal field theory we concisely derive the axion emission rate, applying it to astrophysical and laboratory scenarios. For the Sun, depending on the magnetic field profile plasmon-axion conversion can dominate over Primakoff production at low energies ($\lesssim 200\,$eV). This both provides a new axion source for future helioscopes and, in the event of discovery, would probe the magnetic field structure of the Sun. In the case of white dwarfs (WDs), plasmon-axion conversion provides a pure photon coupling probe of the axion, which may contribute significantly for low-mass WDs. Finally we rederive and confirm the axion absorption rate of the recently proposed plasma haloscopes. 

\end{abstract}

\maketitle

\medskip

\section{Introduction}
The absence of CP violation in the quantum chromodynamics (QCD) sector is still a pressing mystery in Particle Physics. The solution of the Strong CP problem based on the Peccei-Quinn mechanism makes the QCD axion a very well motivated extension of the Standard Model~\cite{Dine:1981rt, Kim:1979if, Peccei:1977hh}. While the QCD axion is a pseudo-Goldstone boson defined by the interaction with gluons through the QCD coupling $a G\tilde{G}$, it also has model independent couplings to electromagnetism and to matter~\cite{Kim:1986ax,Kim:2008hd}.
The QCD axion is also a viable candidate for dark matter (DM)~\cite{Preskill:1982cy,Abbott:1982af,Dine:1982ah,Bergstrom:2000pn,Jaeckel:2010ni,Feng:2010gw}. Inspired by the ``leave no stone unturned'' principle and supported by string theory predictions~\cite{Svrcek:2006yi,Arvanitaki:2009fg,Marsh:2015xka}, axion-like-particles (``axions'' in the rest of this paper) generalize the QCD axion, as their mass is not fixed by the QCD coupling.

In the last three decades an increasingly intense theoretical and experimental effort has been dedicated to the search of such particles~\cite{Turner:1989vc,Raffelt:1996wa,Arias:2012az,Irastorza:2018dyq}. Most of these efforts take advantage of the axion coupling to transverse photons via the electromagnetic tensor, in the context of both astrophysical and laboratory probes. 
We focus on a less explored production (and detection) channel: the coupling between axions and longitudinal plasmons, electromagnetic excitations allowed by the presence of a medium.

The conversion of longitudinal plasmons in the presence of strong magnetic fields has first been pioneered in Ref.~\cite{Mikheev:1998bg} in the context of supernovae, though more recently the topic of axion-plasmon mixing has been revisited~\cite{Das:2004ee,Ganguly:2008kh,Visinelli:2018zif,Mendonca:2019eke}. Inspired by similar works which focused on scalar and vector resonant conversion~\cite{Hardy:2016kme,Redondo:2013lna,Pospelov:2008jk}, we recast the calculation involving a pseudoscalar and an external magnetic field using thermal field theory, applying it to both astrophysical and laboratory systems. While it has already been used once, the approach based on thermal field theory is not widely spread in the literature and has been applied only on the production of axions in the magnetosphere of a magnetar~\cite{Mikheev:2009zz}.

We aim to revitalize plasmon-axion conversion in astrophysical environments and expand the work of Ref.~\cite{Mikheev:2009zz} to new systems. In particular we consider the experimentally relevant systems of the Sun, white dwarfs (WDs), and the recently proposed plasma haloscopes~\cite{Lawson:2019brd}. While for the Sun the total axion luminosity due to the plasmon-axion conversion process is subdominant, this is not true in the low energy regime for the differential flux.
As axion production from plasmon conversion in strong magnetic fields dominates in some energy ranges over more studied processes (such as the Primakoff effect, the conversion of photons in the electric field generated by nuclei~\cite{Raffelt:1985nk}), these results have important implications for building axion observatories, for example motivating $1-100\,$eV-scale helioscopes. In WDs, we identify low mass, highly magnetized WDs as an ideal target. For higher mass WDs, the large core density and correspondingly high plasma frequency prohibits plasmon-axion conversion except in an outer shell. 
%%%%
%%%%%%%%
%%%%%%%%
%%%%%%%%
%%%%%%%%
%%%%%%%%
%%%%%%%%
%%%%%%%%
%%%%%%%%
%%%%
\section{Axion production from a thermal bath of photons}
\label{prodsec}
The effective Lagrangian which describes the coupling between axions and photons reads
\begin{eqnarray}\label{lagrangian}
	\mathcal{L}_{a\gamma}&=-&\frac{1}{4}g_{a\gamma}aF_{\mu\nu}\tilde{F}^{\mu\nu}=-\frac{1}{2}g_{a\gamma}a(\partial_{\mu}A_{\nu})\tilde{F}^{\mu\nu}\nonumber\\
	&\equiv&\frac{1}{2}g_{a\gamma}(\partial_{\mu}a)A_{\nu}\tilde{F}^{\mu\nu}\,,
	\end{eqnarray}
where the effective coupling $g_{a\gamma}$ accounts for both the mixing of axions and pions as well as a field-induced part given by a loop of fermions which couples to both the axion and the photon. Using thermal field theory we can concisely and elegantly derive the axion emission rate, first calculated with other methods in Ref.~\cite{Mikheev:1998bg}. To begin, we recall that the emission rate of a boson by a thermal medium is related to the self-energy of the particle in the medium~\cite{Weldon:1983jn, Kapusta:2006pm}
\begin{equation}
	{\rm Im}\,\Pi=-\omega \Gamma\,,
	\label{weldon}
\end{equation}
where ${\Gamma=\Gamma_{\rm abs}-\Gamma_{\rm prod}}$ is the rate by which the considered particle distributions approach thermal equilibrium. Using the principle of detailed balance the desired thermal production is found to be
\begin{equation}
	\Gamma^{\rm axion}_{\rm prod}=-\frac{{\rm Im}\, \Pi_{\rm{axion}}}{\omega\Big(e^{\omega/T}-1\Big)}\,.
\end{equation}
Therefore, we need to calculate the self-energy of the axion in the medium and then take the imaginary part.

\begin{figure}
\centering
  \vspace{-1.5cm}
 \includegraphics[width=1.\linewidth]{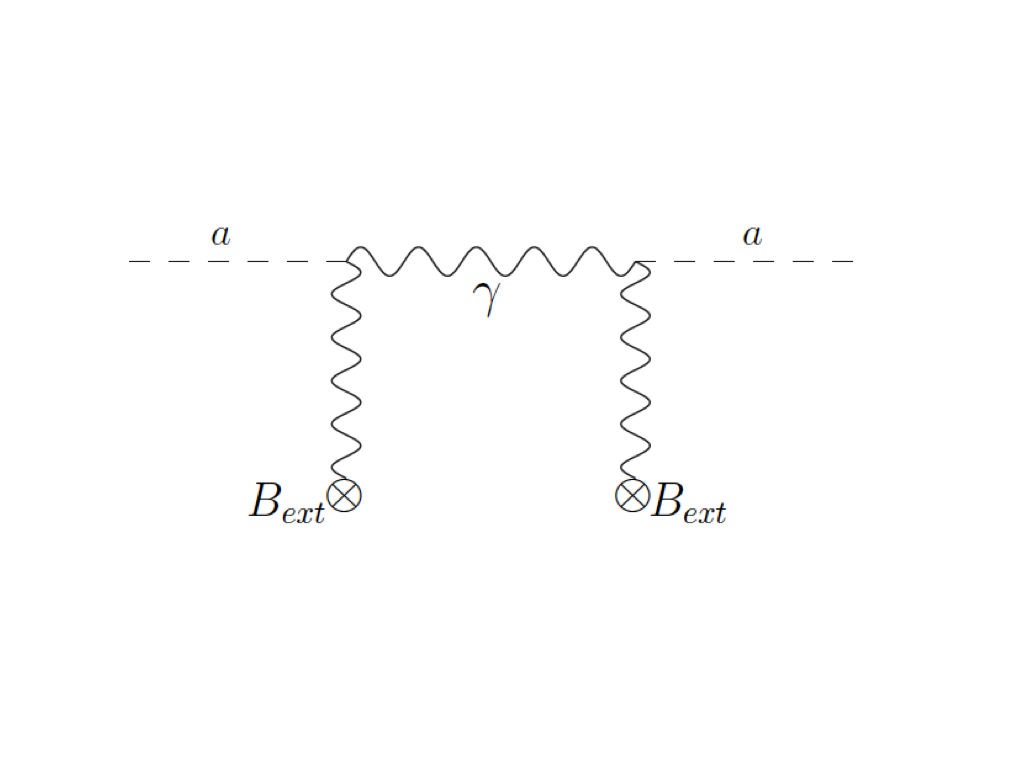}
  \vspace{-2cm}
\caption{\label{fig:self}
Axion self-energy due to interaction with photons in the plasma and the external magnetic field.}
\end{figure}

The axion self-energy due to an external magnetic field at lowest order is depicted in Fig.~\ref{fig:self}. Each vertex brings a factor $m_ag_{a\gamma}B$ and the self-energy is easily written as
\begin{align}\label{selfenergy}
	\Pi_{\rm axion}=m_a^2+m_a^2 g_{a\gamma}^2B_{||}^2\frac{1}{K^2-\Pi_{\gamma,L}(K)}\nonumber\\+m_a^2 g_{a\gamma}^2B_{\perp}^2\frac{1}{K^2-\Pi_{\gamma,T}(K)}\,,
\end{align}
where $K=(\omega,\bf{k})$ is the four-momentum of the external axion, $\Pi_{\gamma}$ is the self-energy of the photon and where we used the basis vector for the longitudinal degree of freedom (in Lorentz gauge)~\cite{Raffelt:1996wa}
\begin{equation}
    \epsilon_l \equiv \frac{(k^2,\omega\bold{k})}{k\sqrt{K^2}}\, .\label{Polarisation}
\end{equation} 
The vertex factors $B_{||}$ and $B_{\perp}$ are the parallel and perpendicular projections of the B-field onto $\hat  {\bf k}$. Thus depending on the projection of the B-field both transverse and longitudinal photon modes contribute to the self-energy.

 Thus the photon self-energy is the quantity we need to complete our computation; the real part in the nonrelativistic approximation is easily found in the literature and, to lowest order in electron velocity, is given by~\cite{Raffelt:1996wa,Redondo:2013lna}
 \begin{subequations}
\begin{alignat}{2}
	{\rm Re}\, \Pi_{\gamma,T}&=\omega_p^2\,,\\
	{\rm Re}\,\Pi_{\gamma,L}&=\frac{K^2}{\omega^2}\omega_p^2\,,
\end{alignat}
\end{subequations}
where $\omega_p$ is the plasma frequency. We see that the dispersion relation for the transverse plasmon gives $\omega^2~-~k^2=~\omega_p^2$, with the usual interpretation of transverse excitations as particles with mass $\omega_p$. The latter is given in the nonrelativistic limit in terms of the electron density $n_e$ by
\begin{equation}\label{nonrer_renor}
    \omega_p^2=\frac{4\pi\alpha n_e}{m_e} \, .
\end{equation}
The longitudinal plasmon, on the other hand, has a peculiar dispersion relation, so that in the nonrelativistic limit $\omega$ is independent from $k$ (see Fig.~\ref{fig:dispersion}).
The imaginary part of the photon self-energy is related, as stated above, to the rate $\Gamma_{\gamma}$ associated to electron-nucleus bremsstrahlung, Compton scattering or other processes keeping photons in thermal equilibrium.\\ For the longitudinal channel we are interested in, we can define the vertex renormalization constant $Z_L$~\cite{An:2013yfc}
\begin{equation}
	K^2=Z_L^{-1}\omega^2\,,
\end{equation}
relevant for the coupling of external photons
or plasmons to electrons in the medium. Working in the static limit one can deduce~\cite{Raffelt:1996wa} that magnetic fields associated with stationary currents are the same at distance, whether or not the plasma
is present. The same is of course not true for the electric field, which gets affected by screening effects.

Neglecting the transverse part one can write
\begin{equation}
	{\rm Im}\,\Pi_{\rm axion}=m_a^2 g_{a\gamma}^2B_{||}^2{\rm Im}\,\frac{Z_L}{\omega^2-\omega_p^2-iZ_L{\rm Im}\,\Pi_{\gamma,L}}\,,
\end{equation}
where the factor $Z_L$ can be interpreted as renormalizing the coupling to the axion. Then, using~Eq.~\eqref{weldon} we can interpret ${-Z_L{\rm Im}\,\Pi_{\gamma,L}/\omega}$ as the damping rate for longitudinal quanta $\Gamma_L$
\begin{equation}
	\Gamma^{\rm axion}_{\rm prod}=\frac{ g_{a\gamma}^2B_{||}^2}{e^{\omega/T}-1}\frac{\omega^2\Gamma_L}{(\omega^2-\omega_p^2)^2+(\omega \Gamma_L)^2}\,.
\end{equation}
Given that $\Gamma_L \ll \omega_p$, we notice a resonance for ${\omega \simeq \omega_p}$, which gives a $\delta$-function peaked around the plasma frequency (we will always need to integrate over phase-space)
\begin{equation}
	\Gamma^{\rm axion}_{\rm prod}\simeq \frac{ g_{a\gamma}^2B_{||}^2}{e^{\omega/T}-1} \frac{\pi}{2}\delta(\omega-\omega_p)\,,
\end{equation}
where we used the definition of the Dirac $\delta$-function
\begin{equation}
	\lim_{\epsilon \rightarrow 0}\frac{\epsilon}{\epsilon^2+x^2}=\pi \delta(x)\,.
\end{equation}
Interestingly, for the resonant production of axions we do not need to calculate a production rate for either axions or photons~\cite{Redondo:2013lna}. 
 \begin{figure}
\centering
  \includegraphics[width=0.9\linewidth]{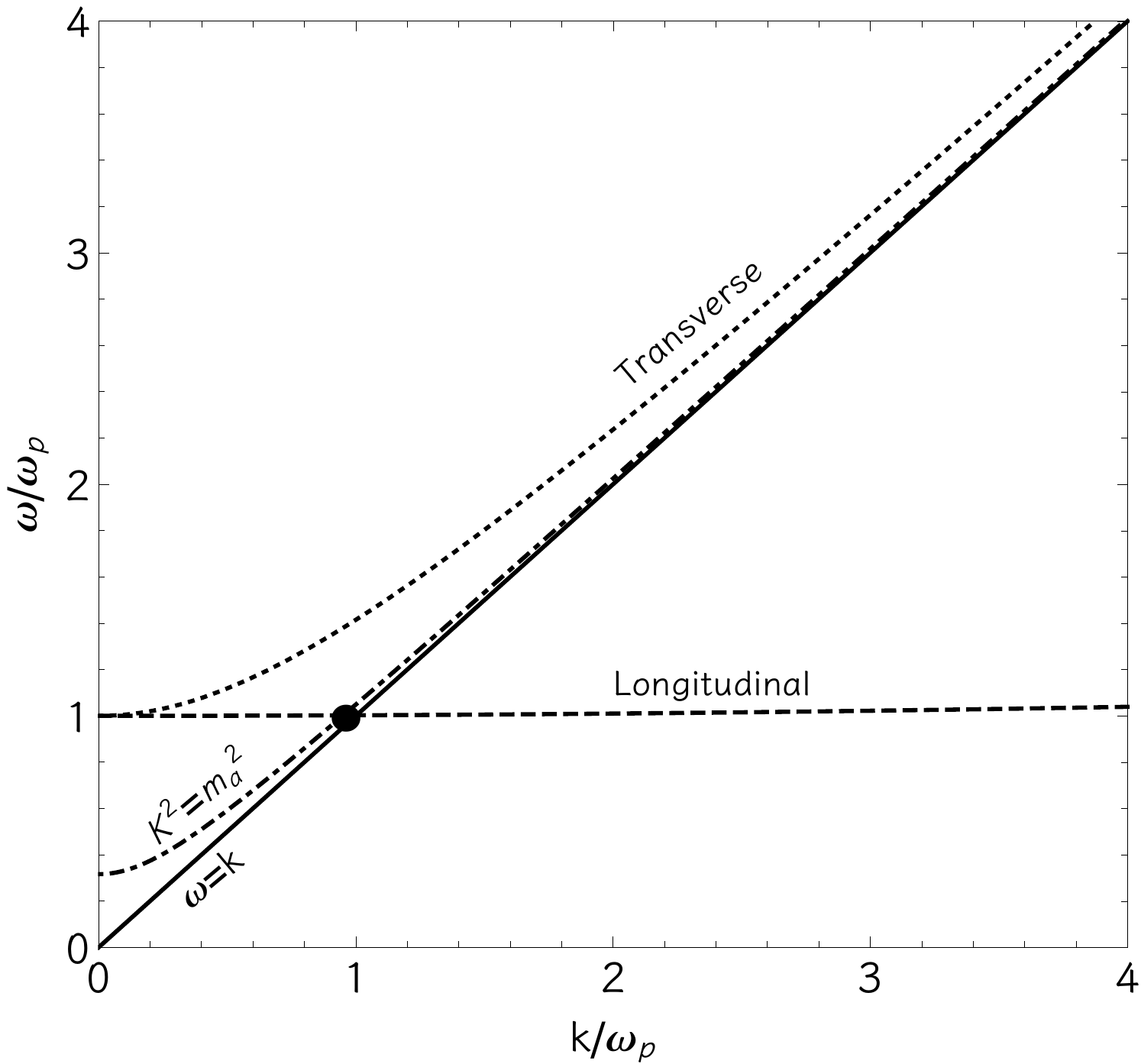}
  \vspace{-.2cm}
\caption{\label{fig:dispersion}
Dispersion relation for an axion with mass $m_a$ (dotted-dashed line), which we make large enough to be distinguishable from that of an ordinary photon (solid line), and the two plasmon modes, longitudinal (dashed) and transverse (dotted). For $m_a\leq \omega_p$ the dispersion relation of a longitudinal plasmon always crosses the dispersion relation of the axion, thus leading to a resonant conversion.}
\label{fig:dispersion}
\end{figure}  

With this result and ${B_{||}=\hat{\bf {k}}\cdot{\bf B}}$ the energy loss due to axion emission reads
\begin{eqnarray}
	Q&=&\int\frac{d^3\bold{k}}{(2\pi)^3}\omega \frac{ g_{a\gamma}^2(\hat{\bf {k}}\cdot{\bf B})^2}{e^{\omega/T}-1} \frac{\pi}{2}\delta(\omega-\omega_p) \nonumber \\ &=&\frac{ g_{a\gamma}^2B^2}{12\pi (e^{\omega_p/T}-1)}\omega_p^3\,,
	\label{poweraxion}
\end{eqnarray}
where we assumed $m_a\ll \omega_p$. Our Eq.~\eqref{poweraxion} agrees with expressions previously derived with a different formalism in Ref.~\cite{Mikheev:1998bg,Mikheev:2009zz}. Note that there is a misprint in Eq.~(1) of Ref.~\cite{Mikheev:1998bg} that gives factor of two relative to our Lagrangian in Eq.~\eqref{lagrangian}.

This is an energy loss per unit volume, meaning we just need to integrate it over the volume of the astrophysical object we consider to obtain the total luminosity. Different objects will have different temperature, plasma frequency and magnetic field, each of those being in principle function of the position inside the star. 
%%%%%%%%%%%%%%%%%%%
%%%%%%%%%%%%%%%%%%%
%%%%%%%%%%%%%%%%%%%
%%%%%%%%%%%%%%%%%%%
\section{Energy loss in stars}
In this Section we apply the results obtained in Section~\ref{prodsec} to two of the most experimentally relevant astrophysical systems, namely the Sun and WDs.

%\begin{widetext}
\begin{figure*}
\hspace*{-1cm}
\centering
  \includegraphics[width=1.05\linewidth]{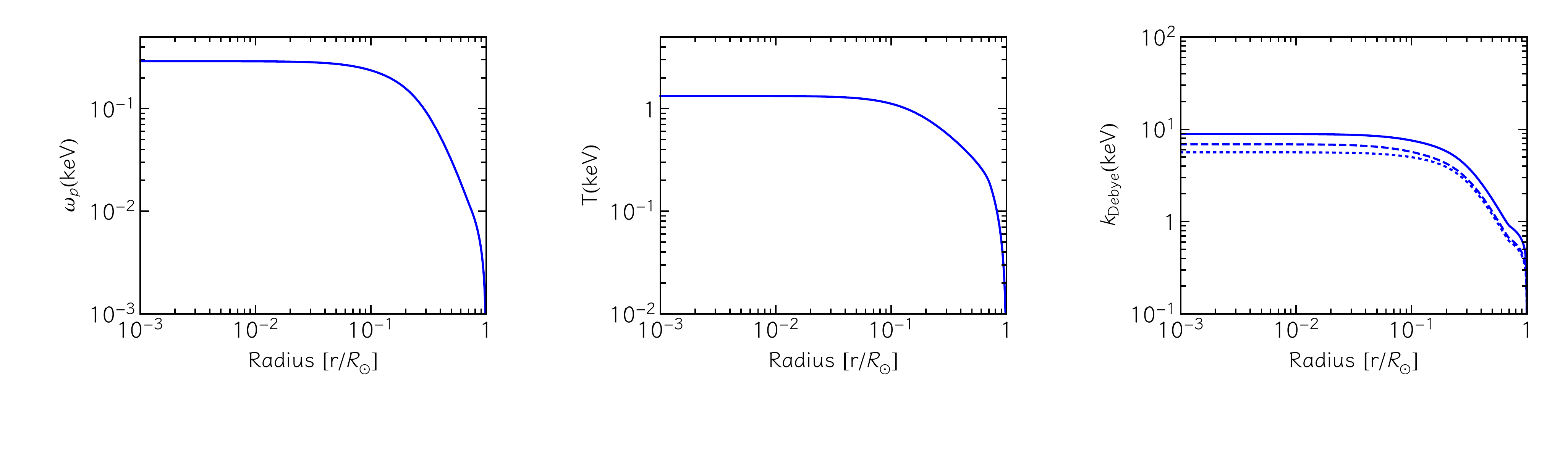}
  \vspace{-1.5cm}
\caption{\label{fig:sunproperties}
Internal solar properties from the reference Saclay model~\cite{Couvidat:2002bs,TurckChieze:2001ye}. We plot the temperature, the plasma frequency and the Debye screening scale. For the latter we plot the total contribution (solid blue line), as well as the individual ones from electrons (dotted blue) and ions (dashed blue).}
\label{fig:sunproperties}
\end{figure*} 
%\end{widetext}

\subsection{Plasmon conversion in the Sun}  
While the energy lost to axions does not have a measurable effect on the Sun, the solar axion flux can be detected by helioscope searches~\cite{Dicus:1978fp,Sikivie:1983ip,Raffelt:1985nk,Raffelt:1987np,Redondo:2013wwa}. In order to calculate the axion luminosity associated with the axion production from plasmon conversion, we need to know the temperature, the plasma frequency and the magnetic field profiles of the star. All these quantities can depend strongly on the radius, and their values are crucial to determine the importance of the process. They are obtained from a solar model, evolving  several  initial  conditions  (mass,  helium  and metal  abundances)  through  a  stellar  evolution  code.  The latter depends  in  turn   on radiative opacities, convection, and so forth, to fit the present-day radius, luminosity, and photospheric composition.  While photospheric composition estimations can vary, the temperature and the plasma frequency profiles of the Sun predicted by different solar models are consistent to a degree sufficient for our purposes. Anticipating that the most relevant effect will be at low energies, we follow Ref.~\cite{Vitagliano:2017odj} and choose the reference Saclay model~\cite{Couvidat:2002bs,TurckChieze:2001ye}, constructed when the surface chemical composition GS98~\cite{Grevesse:1998bj} was suitable to reproduce heliseismology measurements, as it is to our knowledge the most complete concerning external layers. A comparison  with the model of Ref.~\cite{Serenelli:2009yc} with AGSS09 abundances~\cite{Asplund:2009fu} shows that the plasma frequency and the temperature uncertainty stemming from the solar model is less than a few percent, so we the flux produced by processes which depend only on these quantities have theoretical uncertainties of less than 10\%. For the  magnetic field the situation is less clear. In fact, there is no well-established picture of the magnetic field in the interior of the Sun~\cite{Friedland:2002is,2009LRSP....6....4F}; hence we will consider three scenarios:
\begin{enumerate}
	\item The simplest scenario in which the magnetic field is constant over the entire star. We assumed different values for $B$. The range we considered spans from the most pessimistic to the most optimistic case: ${\bar{B}=[10^5,\,7\times10^6] \rm \, G}$~\cite{Friedland:2002is}. In the following we will show  the results for the case ${\bar{B}=10^5 \rm \, G}$;
	\item For a more nuanced model, we parametrize the magnetic field with a step function, taking ${B=7\times10^6}\,$G~\cite{Friedland:2002is} in the interior region of the Sun up to the beginning of the convective zone (${\simeq 0.75\,R_{\odot}}$), where we assume ${B=10^3\rm \, G}$~\cite{2009LRSP....6....4F};
	\item Finally we considered the seismic Solar model of Ref.~\cite{Couvidat:2003ba}, where the authors studied in detail the solar neutrino fluxes and divided the magnetic structure of the Sun in three zones: the  radiative interior, the tachocline and the upper layers. For these three regions different possibilities were considered; as an example we considered here the model of Ref.~\cite{Couvidat:2003ba} named seismic-$B_{21}$.
\end{enumerate}

In Fig.~\ref{sunresult} we show the luminosity 
\begin{equation}
	L_a=\int_{\odot} d^3\bold{r} \frac{ g_{a\gamma}^2B(r)^2}{12\pi (e^{\omega_p(r)/T(r)}-1)}\omega_p(r)^3\,
\end{equation}
as a function of the effective coupling between axions and photons. The solid red curve corresponds to the seismic-$B_{21}$ model of Ref.~\cite{Couvidat:2003ba}. 
The dashed red line corresponds to the luminosity from plasmon conversion with the magnetic field configuration 
\begin{equation}
    B(r)=[7\times10^6\,\theta(0.75R_{\odot}-r)+10^3\,\theta(R_{\odot}-r)] \rm \ G \, ,
\end{equation} where $\theta$ is the Heaviside step function. Finally the dotted red lines was drawn assuming a constant magnetic field of $\bar{B}=10^5\,\rm G$.

We notice that in all the following computations we assume the plasma frequency to be given by the free electron contribution only. Species which are not completely ionized could contribute significantly to the plasma frequency~\cite{Redondo:2015iea}, but we expect the error in neglecting the contribution of the bound-bound transition to be comparable or smaller than the uncertainty on the magnetic field in the interior of the Sun. This effect would be most significant in the external layers of the Sun.

To compare with the Primakoff effect, we estimate the total energy loss~\cite{Raffelt:1996wa} 
  \begin{equation}
  	Q = \frac{g_{a\gamma}^2T^7}{4 \pi}F({k_{\rm S} }^2)\,,
  \end{equation}
  where
    \begin{equation}
  	k_{\rm S}^2 = \frac{4\pi \alpha}{T}n_e+\frac{4\pi \alpha}{T}\sum_j n_j Z_j^2\,
  \end{equation}
  is the Debye screening scale in a nonrelativistic, nondegenerate plasma and
  \begin{align}
  	F(k_{\rm S}^2)&=\frac{k_{\rm S}^2}{2\pi^2T^6}\int_0^\infty d\omega\, \frac{\omega}{e^{\omega/T}-1}\nonumber \\
  	&\times\left[ (\omega^2+k_{\rm S}^2)\ln\left(1+\frac{\omega^2}{k_{\rm S}^2}\right)-\omega^2\right] \,. \label{eq:F}
  \end{align}
  The same expression can be found in the framework of thermal field theory~\cite{Altherr:1992mf,Altherr:1993zd}.
  We stress that while ions do not contribute to the plasma frequency, as forward scattering is suppressed by their large mass, they should be included when estimating the Debye screening scale. 
  The Primakoff contribution is shown as a dashed blue line; unless the magnetic field takes unrealistic large values, the plasmon conversion is a negligible correction to the luminosity generated by the Primakoff effect. However, this does not mean that the Primakoff effect is the dominant source of axions all over the spectrum. 
\begin{figure}
\centering
  \includegraphics[width=1.0\linewidth]{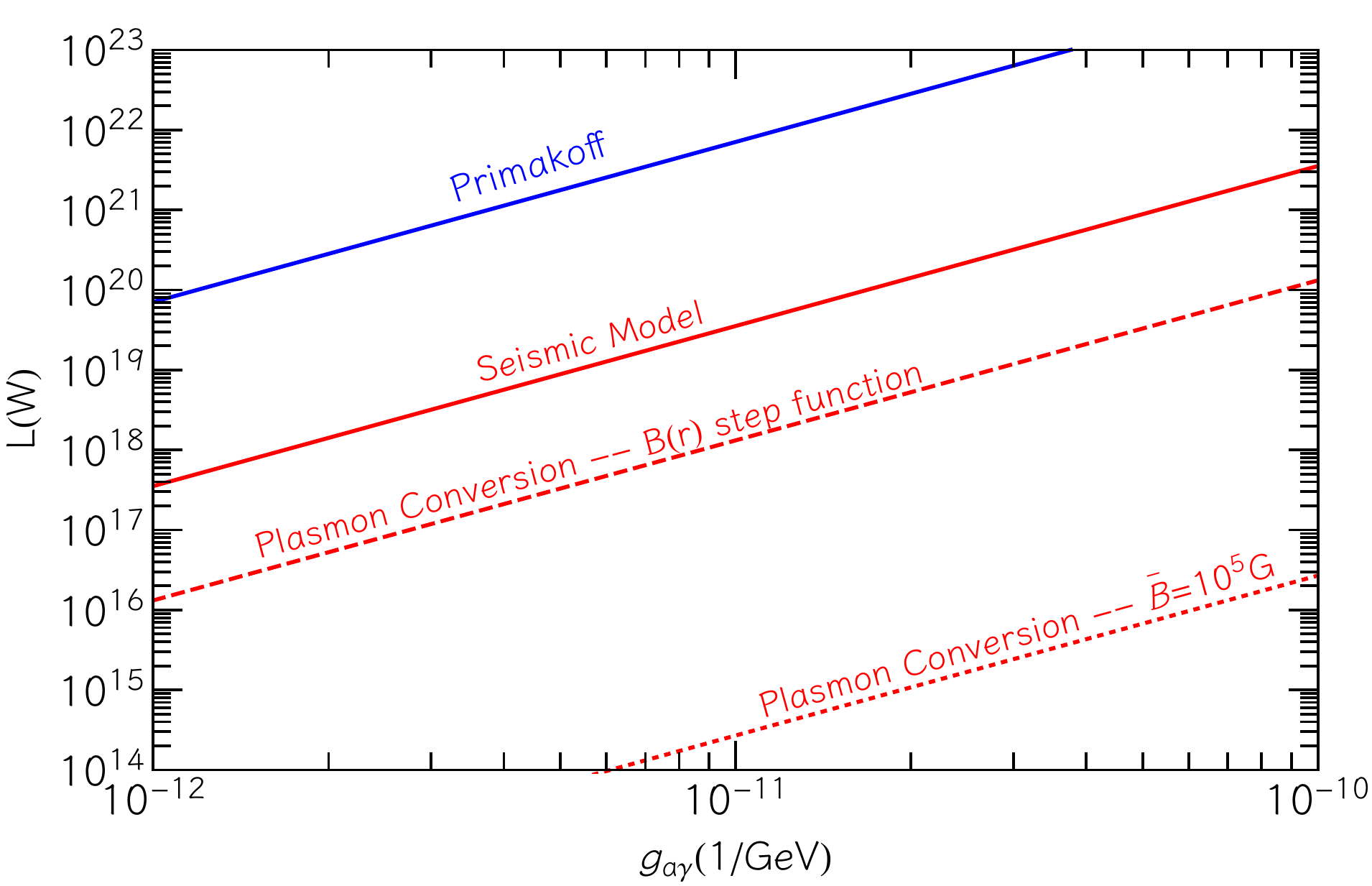}
  \vspace{-.1cm}
\caption{\label{sunresult}
Luminosity associated with the Primakoff effect (solid blue line) and plasmon conversion in the interior of the Sun (red curves). For the latter, the solid line corresponds to the seismic-$B_{21}$ model of Ref.~\cite{Couvidat:2003ba}, the dotted red line corresponds to ${B(r)=\bar{B}=10^5}\,$G and the dashed red line corresponds to \hbox{$B(r)=7\times10^6\,\theta(0.75R_{\odot}-r)+10^3\,\theta(R_{\odot}-r)\,$G}. }
\label{sunresult}
\end{figure}

As the energy dependence is different between plasmon conversion and the Primakoff effect, it is worth to investigate the differential axion flux to the Earth. For the Primakoff effect the differential flux at Earth is usually expressed by~\cite{Andriamonje:2007ew,Raffelt:1996wa}
\begin{equation}
	\frac{d\Phi_{\rm Pr}}{d\omega}=g_{10}^2\, 6.02\times 10^{10}{\rm cm}^{-2}{\rm s}^{-1}{\rm keV}^{-1}\frac{(\omega/{\rm keV})^{2.481}}{e^{(\omega/{\rm keV})/1.205\,}}\,,
	\label{primakoff}
\end{equation}
However, while this is a good approximation in the range $[1-11] \, \rm keV$, it is not accurate for the low energy tail we are interested in. We therefore computed the differential flux using our solar model of reference~\cite{Couvidat:2002bs,TurckChieze:2001ye} and the axion emission rate~\cite{Raffelt:1996wa,Raffelt:1987np, Jaeckel:2006xm}
\begin{equation}
    \Gamma_{\gamma\rightarrow a}=\frac{g_{a\gamma}^2k_{\rm S}^2T}{64\pi}\int^{1}_{-1}d\cos\theta\frac{\sin^2\theta}{(x-\cos\theta)(y-\cos\theta)}\,,
\end{equation}
where $x=(k_a^2+k_\gamma^2)/2k_ak_\gamma$ and $y=x+k_s^2/2k_ak_\gamma$.
The dispersion relation of the photon requires the frequency of the photon to be always larger than the plasma frequency at a given radius. This requirement further suppresses Primakoff contribution at low frequencies, as production only occurs in the outer layers of the sun. We show the Primakoff differential flux in Fig.~\ref{fig:unified} (solid blue curve), while in the Appendix.~\ref{appendix: primakoff} we give some details of the calculation. 
%In fact we find that Eq.~\ref{primakoff} overestimates the differential flux when extrapolated to low energies, in particular by a factor of $\sim 20$ at $\omega = 10^{-3} \rm eV$.

The flux produced by the longitudinal plasmon conversion reads instead
\begin{align}
	\frac{d\Phi_{\rm pl}}{d\omega}=\frac{1}{4\pi (1\rm \, AU)^2}\int_{\odot} d^3\bold{r} \frac{\omega^2}{(2\pi)^3} \frac{ g_{a\gamma}^2B^2}{e^{\omega/T}-1} \frac{2\pi^2}{3}\delta(\omega-\omega_p)\nonumber\\ =\frac{1}{12\pi (1\rm \, AU)^2}\int_0^{R_{\odot}} dr\,r^2 \frac{\omega^2g_{a\gamma}^2B(r)^2}{e^{\omega/T(r)}-1}\delta(\omega-\omega_{p}(r))\,,
\end{align}
where $\delta$-function will be used to integrate over the radius. For a given axion frequency $\omega$ the equality ${\omega=\omega_p(r_0)}$ fixes the value of the radius $r_0$ at which the integrand needs to be evaluated. Therefore
\begin{equation}
	\frac{d\Phi_{\rm pl}}{d\omega}=\frac{1}{12\pi (1\rm \, AU)^2}r_0^2\,\frac{\omega^2g_{a\gamma}^2 B(r_0)^2}{e^{\omega/T(r_0)}-1} \frac{1}{|\omega_p'(r_0)|}\,,
\label{plasmon}
\end{equation} 
which we notice to have a different functional dependence on the energy with respect to Eq.~\eqref{primakoff}. 
Compared to the Primakoff process, which produces a peak in the spectrum around $3-4\,$keV~\cite{Raffelt:1987yu,Raffelt:1996wa,1989PhRvD..39.2089V}, axion-plasmon conversion has peaks at very low frequency, $\simeq 1-100\,$eV depending on the assumed magnetic field. 
  This shift is due to the fact that the axion frequency matches the plasma frequency, which is limited to relatively small values, $\omega_p^{\odot} \lesssim 0.3\,$keV.
  
  In Fig.~\ref{fluxes} we show the differential axion flux from longitudinal plasmon conversion, which overcomes Primakoff conversion at low frequencies ($\lesssim 200\,$eV).
  %while the Primakoff conversion starts rapidly to win at larger frequencies where the plasmon conversion is ineffective. In Fig.~\ref{fluxes}
  As for the luminosity in Fig.~\ref{sunresult}, we show the results for three different configurations of the the internal magnetic field. Interestingly enough the spectral features of axion emission map the magnetic structure of the sun. Consider the seismic  model (solid red curves): three regions are evident from Fig.~\ref{fluxes}, which correspond to different shells of the sun and consequently to different plasma frequencies (hence axion energies) and different magnetic fields. For example, the radiative interior corresponds in Fig.~\ref{fluxes} to the solid red curve at higher energies. In this region  the density and the plasma frequency are high, therefore the produced axion will have large energies $\omega \gtrsim 2 \times 10^{-2} \,\rm KeV$; furthermore the magnetic field is $5 \times 10^{7}\,\rm G$, which significantly enhances the conversion rate. 

\begin{figure}
\hspace{-0.3cm}
\centering
  \includegraphics[width=0.95\linewidth]{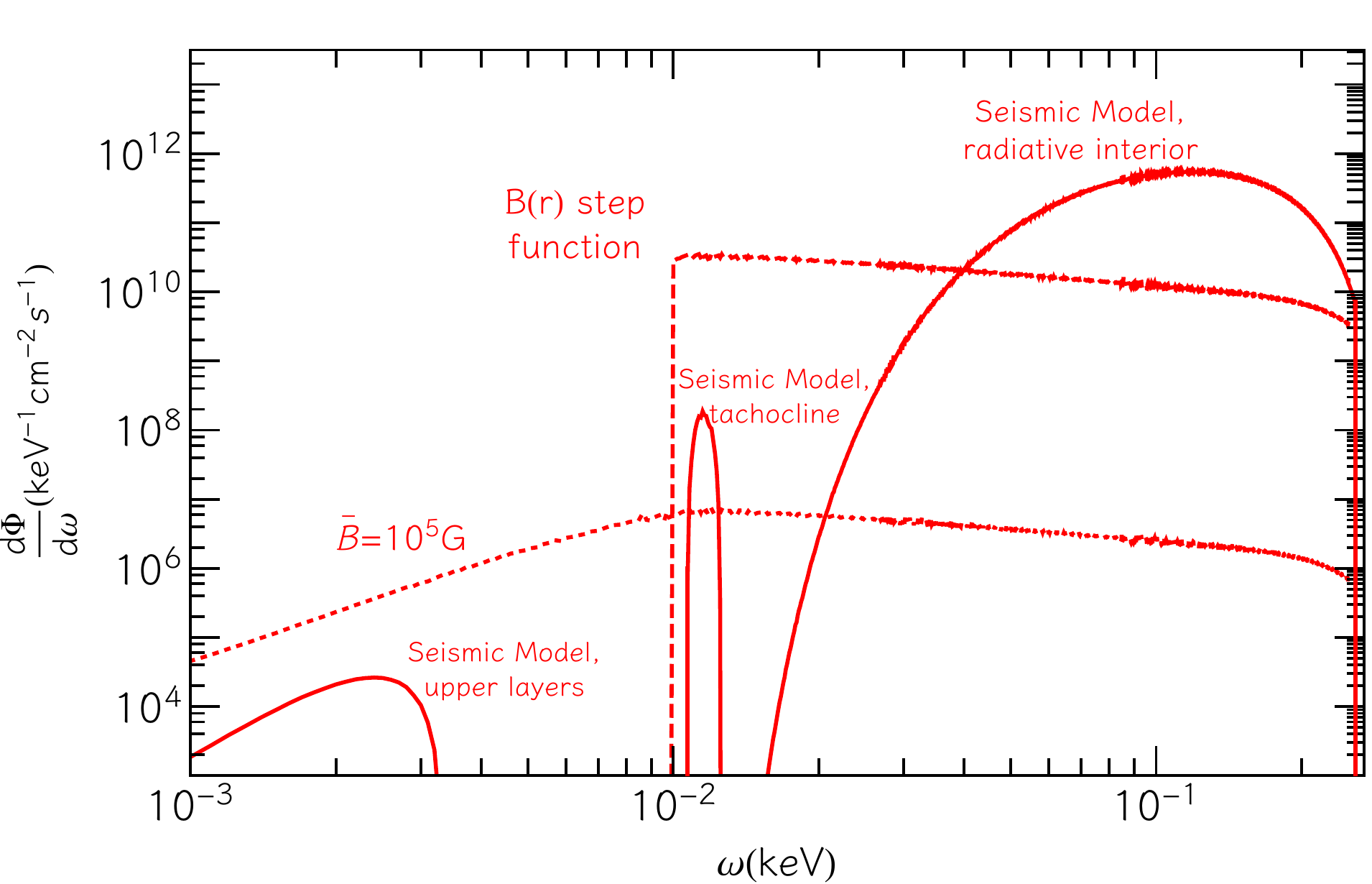}
  \vspace{-.1cm}
\caption{\label{fluxes}Differential axion flux on Earth from longitudinal plasmon conversion. We show the flux for three different magnetic field profiles, a step function \hbox{$B(r)=7\times10^6\,\theta(0.75R_{\odot}-r)+10^3\,\theta(R_{\odot}-r)\,$G} (dashed), a constant magnetic field field ${B(r)=\bar{B}=10^5}\,$G (dotted) and the seismic-$B_{21}$ model (solid).}
\label{fluxes}
\end{figure}

The lower energy peaks of plasmon-axion conversion are potentially very interesting for axion helioscope designs. Helioscope such as CAST or IAXO~\cite{Anastassopoulos:2017ftl,Armengaud:2014gea} are designed for X-ray energies, where cavities and optics are very difficult to build. While the flux in the high UV is lower, it may prove to be more easily instrumented, or be enhanced by a mildly resonant cavity. We plot the axion spectrum for $g_{a\gamma}=10^{-10}\,{\rm GeV}^{-1}$ from $0.1\,$eV to $10\,$keV in Fig.~\ref{fig:unified}. Here we assume that the axion can always be considered ultrarelativisitc. At the lowest energies, $\lesssim1\,\rm eV$, axions generated by Bremsstralung in the Earth dominate even if the axion electron coupling is suppressed~\cite{Davoudiasl:2009fe}. In this case the differential flux is found to be thermal $d\Phi/d\omega \propto \omega^3/(\exp(\omega/T_c)-1)$, with $T_c$ being the temperature of the Earth's core~\cite{Nakagawa:1988rhp}. We stress that this flux, often overlooked in previous plots of the ``grand unified axion spectrum" (in analogy to photon~\cite{Ressell:1989rz} and neutrino~\cite{Vitagliano:2019yzm} spectra), should fill the gap between axions from the Sun and a population of thermally produced axions constituting dark radiation~\cite{Irastorza:2018dyq}. For the comparison we fixed the axion-electron coupling to be $g_{ae} = 2 \times 10^{-4} |g_{a\gamma}|\,$GeV; this value can vary and may be larger depending on the considered model (see for example the recent review~\cite{DiLuzio:2020wdo}). Here we considered a benchmark value for models where the axion-electron coupling is suppressed, and only occurs at one loop due to the presence of an axion-photon coupling. While relevant for eV scale experiments, the flux produced from the Earth has very different directionalty, meaning that it is experimentally distinct from longitudinal plasmon conversion in the Sun. At high frequencies $\gtrsim 200\,$eV Primakoff production takes over, giving the traditional window for helioscopes. However, while the exact spectrum depends heavily on the magnetic field structure of the sun, longitudinal conversion is very important for intermediate energies. Here we have plotted the seismic-$B_{21}$ model, though this statement holds more generally.

\begin{figure*}
\centering
  \includegraphics[width=0.65\linewidth]{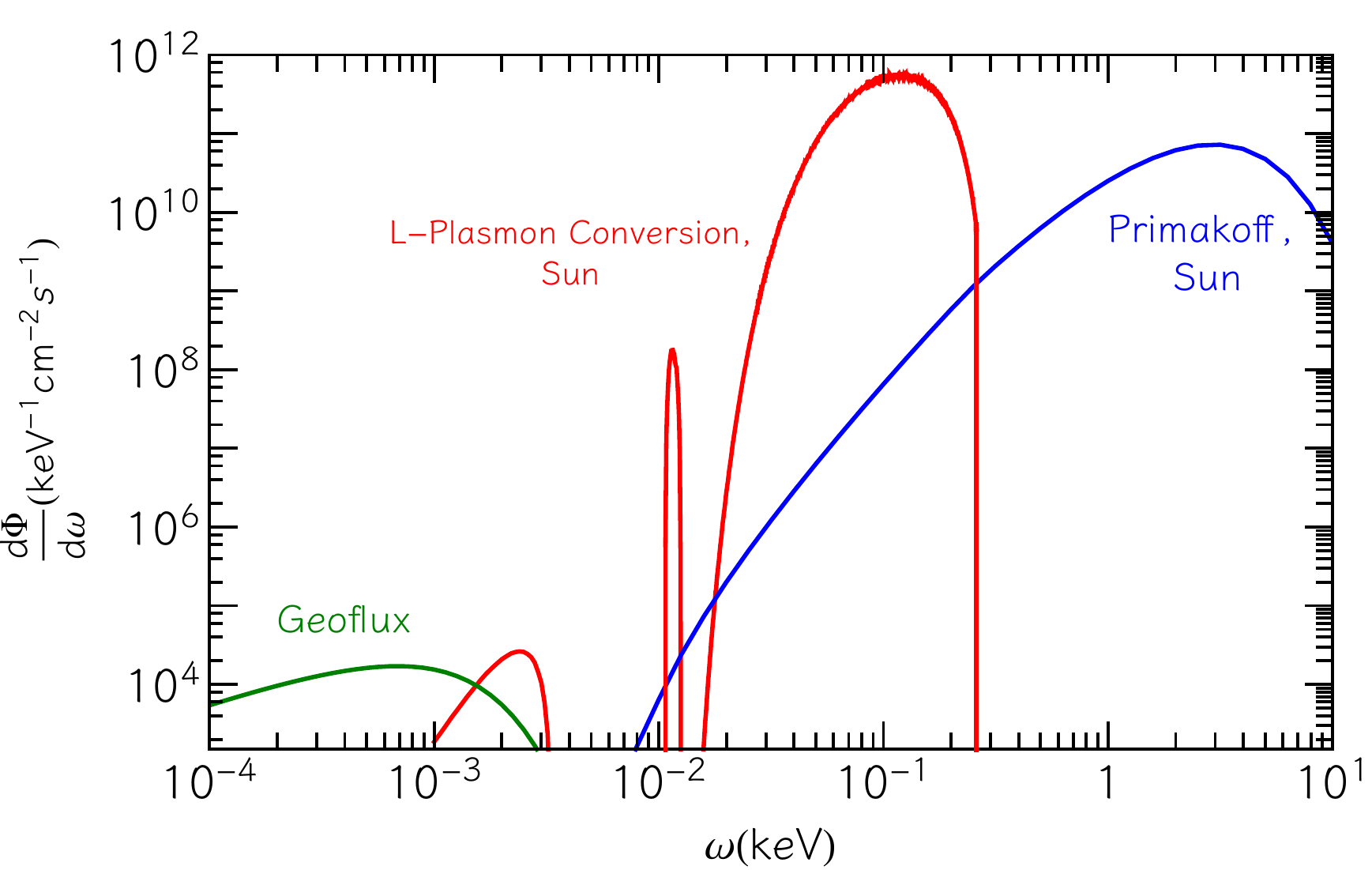}
  \vspace{-.1cm}
\caption{\label{fig:unified}
Differential axion flux ($\rm cm^{-2} \rm s^{-1}\rm keV^{-1}$) in the entire range $0.1\,\rm eV - 10 \rm \, keV$ with $g_{a\gamma}=10^{-10}\,{\rm GeV}^{-1}$. We report the axion flux from Earth (dark green) given in Ref.~\cite{Davoudiasl:2009fe}, and both the Primakoff (blue) and longitudinal plasmon (red) spectra from the Sun as considered in this work. For the longitudinal plasmon conversion we only report the results for the seismic-$B_{21}$ model. As the focus of this work is on the photon coupling, we plot the results of Ref.~\cite{Davoudiasl:2009fe} assuming a suppressed $g_{ae}=2\times10^{-4} |g_{a\gamma}|{\rm GeV}$. }
\label{fig:unified}
\end{figure*}

Intriguingly, emission at this intermediate energy range would allow one to corroborate a signal by showing structure (a double peak in the simplest model) in the axion flux, and give further information on the internal structure of the Sun, as also shown in the context of neutrinos and of axions with electron coupling~\cite{Vitagliano:2017odj, Redondo:2013wwa}. It is evident from Fig.~\ref{fluxes} that at least an upper bound of the Sun's internal magnetic field can be obtained. In fact, as the spectrum produced by plasmons essentially maps the magnetic field as a function of plasma frequency, one should be able to largely reconstruct the internal magnetic field as a function of radius inside the Sun, further showing the potential of ``axion astronomy'' to investigate the interior of the Sun~\cite{Jaeckel:2019xpa}.

\subsection{White Dwarfs}
Now we turn our attention to another interesting astrophysical candidate, WDs. These stars are in some ways a simpler system, being nearly isothermal and degenerate, with typical densities of order ${\rho_{\rm WD}\simeq 1.8 \times 10^6\, {\rm g \,cm}^{-3}}$ and temperatures ${T_{\rm WD} \simeq 3\times 10^6 - 2 \times 10^7 \, {\rm K}}$~\cite{Raffelt:1996wa}. Moreover they often exhibit very strong magnetic fields,
\begin{equation}
	B \gtrsim 100\,\rm MG\,.
\end{equation}
Before proceeding we have to adapt the previous formalism to the degenerate case. 

We will use expressions depending on the temperature $T$ and the electron chemical potential $\mu$ valid in any plasma condition~\cite{Raffelt:1996wa, Braaten:1993jw}. First, we introduce the sum of the phase space distributions for $e^{\pm}$,
\begin{equation}
f_p=\frac{1}{e^{(E-\mu)/T}-1}+\frac{1}{e^{(E+\mu)/T}-1}\,,
\end{equation}
so that the plasma frequency $\omega_p$ and the characteristic frequency $\omega_1$ are given by
  \begin{subequations}
\begin{alignat}{2}
    \omega_p^2 &\equiv \frac{4\alpha_{em}}{\pi}\int^{\infty}_0 dp f_p p(v - 1/3 v^3)\, ,
    \\
    \omega_1^2 &\equiv \frac{4\alpha_{em}}{\pi}\int^{\infty}_0 dp f_p p(5/3 v^3 - v^5)\,,
\end{alignat}
\end{subequations}
where $\alpha_{em}=e^2/4\pi$ and $v = p/E$.
Finally, the real part of the photon self-energy can be written as
\begin{equation}
	{\rm Re}\,\Pi^L_{\gamma}(k)=\omega_p^2(1-G(v_*^2k^2/\omega^2))+v_*^2k^2 - k^2\,,
\end{equation}
where $v_*$ is the ``typical" electron velocity in the medium defined as the ratio $v_* \equiv \omega_1/\omega_p$ and  $G$ is an auxilary function, defined as
\begin{equation}
  G(x)=\frac{3}{x}\left[1-\frac{2x}{3}-\frac{1-x}{2\sqrt{x}}\ln{\left(\frac{1+\sqrt{x}}{1-\sqrt{x}}\right)}\right]\, .
\end{equation}
The imaginary part of the axion self-energy now reads
\begin{align}
	{\rm Im}\,\Pi_{\rm axion}&=m_a^2B_{||}^2g_{a\gamma}^2{\rm Im}\Big(\frac{1}{(K^2-\rm Re \Pi^L_{\gamma})-i{\rm Im}\,\Pi^L_{\gamma}}\Big)\nonumber\\&= m_a^2B_{||}^2g_{a\gamma}^2 \frac{{\rm Im}\, \Pi^L_{\gamma}}{(K^2-\rm Re \Pi^L_{\gamma} )^2+ ({\rm Im}\,\Pi^L_{\gamma})^2}\,,
\end{align}
Interpreting again ${{\rm Im}\,\Pi^L_{\gamma}=-\omega\Gamma_L}$, we are left with
\begin{equation}
	{\rm Im}\,\Pi_{\rm axion} = -m_a^2B_{||}^2g_{a\gamma}^2 \frac{\omega\Gamma^L }{(K^2-\rm Re \Pi^L_{\gamma} )^2+ (\omega\Gamma^L_{\gamma})^2} \,.
\end{equation}
As before we can take the limit of small damping rate to find
\begin{equation}
	{\rm Im}\,\Pi_{\rm axion}=-m_a^2B_{||}^2g_{a\gamma}^2\frac{\pi}{2\omega}Z_L\delta(\omega-\omega_0(k))\,,
\end{equation}
where we have defined the renormalization factor 
\begin{equation}
Z_L = \Big(1 - \frac{\partial \rm Re\Pi^L_{\gamma}}{\partial \omega^2} \Big)^{-1}\,,
\end{equation}
which generalizes Eq.~\eqref{nonrer_renor} to a degenerate plasma~\cite{Raffelt:1996wa}
\begin{equation}
	Z_L=\frac{\omega^2}{\omega^2-k^2}\frac{2(\omega^2-v_*^2k^2)}{3\omega_p^2-(\omega^2-v_*^2k^2)}\,;
\end{equation}
moreover, we defined
\begin{equation}
	\omega_0(k)\equiv \sqrt{\rm Re \Pi^L_{\gamma} + k^2} = \sqrt{(\omega_p^2(1-G(v_*^2k^2/\omega^2))+v_*^2k^2)}\,.
\end{equation}
The emission rate will thus be
\begin{equation}
	Q=\int\frac{d^3\bold{k}}{(2\pi)^3}\frac{\pi g_{a\gamma}^2(\hat{{\bf k}}\cdot {\bf B})^2}{e^{\omega/T}-1}\frac{2\omega(\omega^2-v_*^2k^2)}{3\omega_p^2-(\omega^2-v_*^2k^2)}\delta(\omega-\omega_0(k))\,.
\end{equation}

\begin{figure*}
\centering
  \includegraphics[width=1\linewidth]{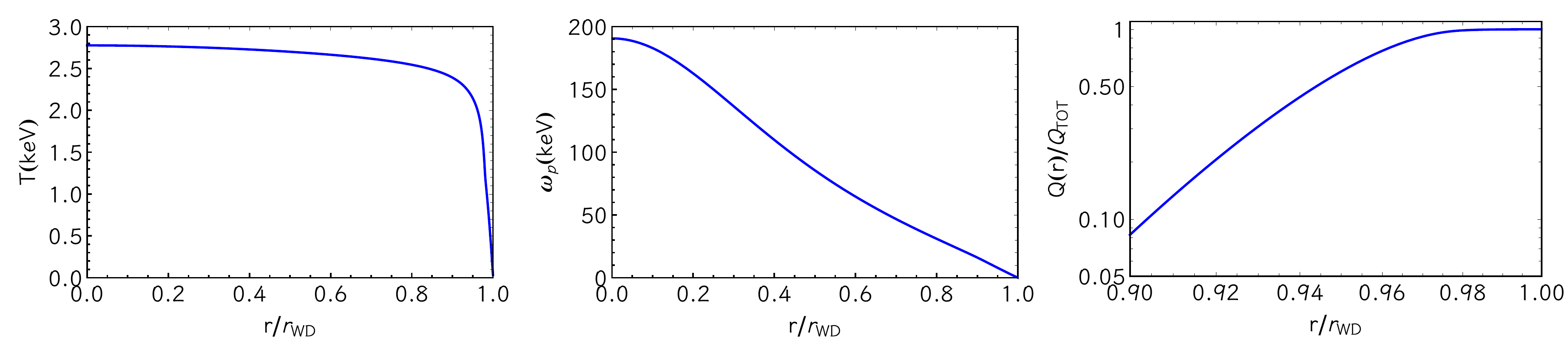}
  \vspace{-0.2cm}
\caption{\label{fig:propertiesWD}
Internal properties of \textbf{Model-1} for the analyzed WD. We plot the temperature (left panel), the plasma frequency (middle panel) as well as the fractional axion flux produced inside a radius $r$ (right panel). It is clear from the right panel that the internal shells of the WD do not contribute to the axion flux, as in those regions the plasma frequency is much larger then temperature, suppressing plasmon population.}
\label{propertiesWD}
\end{figure*}

Degenerate stellar systems have long been used as probes of axions by studying the possibility of energy loss from axion emission~\cite{Raffelt:1985nj,Corsico:2019nmr}. Interestingly, several hints have been measured of a preference for an additional, unaccounted for, cooling channel of these stars~\cite{Giannotti:2015kwo,Giannotti:2017hny}. The observations include the rates of period changes of several systems~\cite{Althaus:2005jt,Corsico:2012ki,Corsico:2012sh,Corsico:2016okh,Battich:2016htm} and the luminosity function, the number of WDs per unit bolometric magnitude and unit volume, which tracks the cooling of these stars~\cite{Bertolami:2014wua}. Other degenerate systems showing excessive cooling are the red giants~\cite{Raffelt:1994ry,Viaux:2013lha}.
More recently the authors of Ref.~\cite{Dessert:2019sgw} considered X-ray signatures of axion conversion in magnetic WD stars: the axions are produced in the interior of the star via the coupling to electrons
\begin{equation}
    \mathcal{L}_{ae} = \frac{g_{ae}}{2 m_e}\bar{e}\gamma^{\mu}\gamma^5 e \, \partial_{\mu}a\,,
\end{equation}
and then converted to photons in the external magnetic field. 
 The main process in this case is electron bremsstrahlung in electron-nuclei scattering. 
 The luminosity associated to this process can be written as as~\cite{Raffelt:1987np,Nakagawa:1988rhp,Bertolami:2014wua}
\begin{equation}\label{luminWD}
	\frac{L_a}{L_{\odot}} \simeq 1.6 \times 10^{-4}\Big(\frac{g_{ae}}{10^{-13}}\Big)^2\Big(\frac{M_{\rm WD}}{1\, M_{\odot}}\Big)\Big(\frac{T}{10^7 \, {\rm K}}\Big)^4\,F \, ,
\end{equation} 
where F is a factor that depends on the density and composition of the star, but is usually $\sim O(1)$. Similarly to the axion geoflux, the differential axion flux is thermal $d\Phi/d\omega \propto \omega^3/(\exp(\omega/T_c)-1)$ with $T_c$ now being the temperature of the WD's core.

Here we want to consider the possibility of breaking the degeneracy between the couplings $g_{a e}$ and $g_{a\gamma}$, producing the axion via the longitudinal plasmon conversion which relies only on the axion-photon coupling. 
Then, once the axions are produced, they travel from the WD center outwards, where they can be converted into photons in the magnetic field surrounding the star~\cite{Raffelt:1987im, Hook:2018iia, Pshirkov:2007st, Dessert:2019sgw, Huang:2018lxq} with a probability $p_{a\rightarrow \gamma}$, which depends mainly on the magnetic field and the coupling $g_{a\gamma}$ (see Appendix~\ref{Appendix: conversion}).
  
 The electromagnetic flux at the Earth reads finally
  \begin{equation}
  	\frac{dF_{\gamma a}}{d\omega}(\omega)=\frac{dL_a}{d\omega} \frac{1}{4\pi d^2_{\rm WD}}p_{a \rightarrow \gamma} \,,
  \end{equation}
  where $d_{\rm WD}$ is the distance  of the WD. 
  While the axion luminosity produced by bremsstrahlung can be expressed with Eq.~\eqref{luminWD}, to compute the contribution of plasmon conversion we will use a detailed WD model. We do this because the axion production depends very strongly on the plasma frequency, which varies strongly as a function of radius, as shown in the middle panel of Fig.~\ref{fig:propertiesWD}. This model is built from asteroseismological observations~\cite{fontaine,Althaus:2010pi,Romero:2011np,Althaus:2003ta,Corsico:2019nmr}.
  
  The model we will employ,\footnote{Leandro G. Althaus and Alejandro H. C\'orsico, private communication.} hereafter referred to as \textbf{Model-1}, describes a WD with mass $M=1.2919 \, M_\odot$ at effective surface temperature $T_{\rm eff}=~40,\!960\,\rm K$. While WDs are isothermal for large part of their profile due to the large electron degeneracy, which implies a long mean free path the electrons and consequently a large thermal conductivity, the Fermi energy (and correspondingly the electron density) will depend on the radius. We thus anticipate that the flux will be larger around $\sim 5$\,keV, as the axions will be produced in the outer shell of the star where the ratio between the plasma frequency and the temperature is smaller, softening the suppression due to the plasmon population (see Fig.~\ref{fig:propertiesWD}). The vast majority of axion-plasmon conversion occurs in the outer $\sim 10\%$ of the star.

 We can now compute the total flux for the longitudinal plasmon conversion and compare it with that from bremsstrahlung using Eq.~\ref{luminWD}. In Fig.~\ref{ratio} we show the comparison between the two effects as a function of the magnetic field. We consider both \textbf{Model-1} (solid line) and a more optimistic scenario \textbf{Model-2} (dashed line) in which we re-scaled the central density to lower values $\rho_{{\rm core}}\sim 10^6 \,{\rm g/cm}^3$. Again, for comparative purposes we fixed the axion-electron coupling to be $g_{ae}=2\times10^{-4}|g_{a\gamma}|\,{\rm GeV}$.
 
 From Fig.~\ref{ratio} it is evident that there is a strong dependence over both the magnetic field and the density profile. In fact, keeping fixed the temperature profile, the smaller the density, the smaller the ratio between the plasma frequency and the temperature. This then implies a thicker shell in which plasmon production is effective. Further, lower density is typically reached in less massive WDs, which are actually larger in radius~\cite{Shapiro:1983du}; for \textbf{Model-2} we thus considered a re-scaled mass of $0.4 \, M_{\odot}$. 
 
 Furthermore, the closer you go to the core of the star the higher is the magnetic field, which is driving the resonant conversion. Indeed some theoretical studies consider the possibility that internal magnetic fields may even be as large as $10^{12}-10^{13}\,$G~\cite{Das:2012ai,Franzon:2015gda}. Theoretical models of WDs with very high core magnetic fields, but low surface magnetic fields, were made in Ref.~\cite{1968ApJ...153..797O}. We thus identify as an ideal target a strongly magnetic WD with small mass and central density. While rare, it may be possible for plasmon conversion to dominate in such a WD.
 In such a situation the observational strategy would be the same as the one discussed in Ref.~\cite{Dessert:2019sgw}, as the signal is an X-ray spectrum with the peak around $5\,$keV (see also Ref.~\cite{Ng:2019gch, Roach:2019ctw} for other possible X-ray analyses).

\begin{figure}
\centering
  \includegraphics[width=0.95\linewidth]{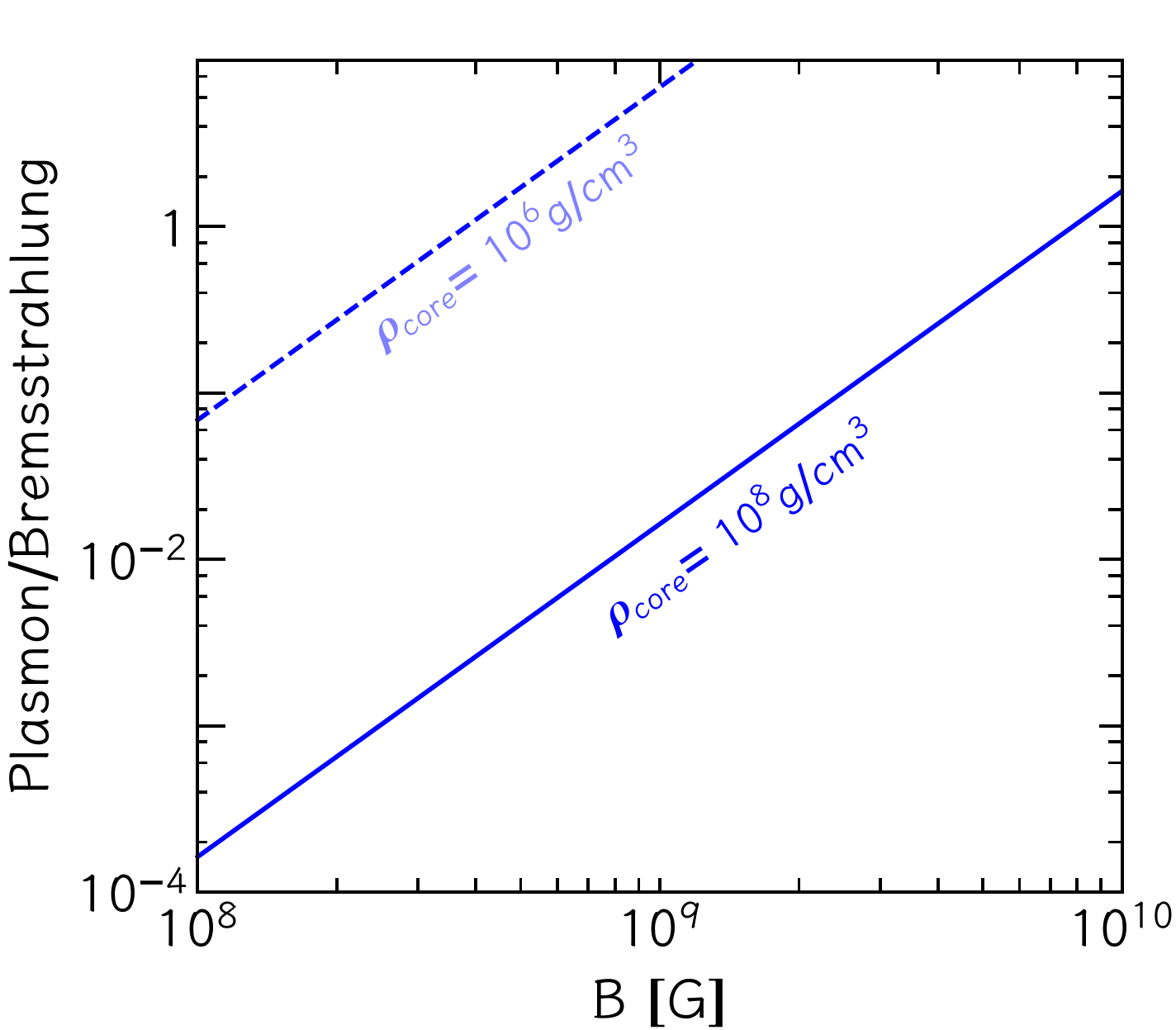}
  \vspace{-.4cm}
\caption{\label{ratio}Ratio of the luminosity due to plasmon conversion and bremsstrahlung as a function of the magnetic field. We show the results for \textbf{Model-1} (solide line) and \textbf{Model-2} (dashed line). The electron coupling has been fixed to $g_{ae} = 2\times 10^{-4} g_{a\gamma} \, {\rm GeV}$.}
\label{ratio}
\end{figure}

  \section{Plasma haloscopes}\label{plasmahaloscope}
Lastly, we turn our attention away from the stars and into the lab. A recent use of plasmas in the literature is the proposal of a cryogenic plasma inside a strong external magnetic field to DM axions on Earth~\cite{Lawson:2019brd}. Such a device would be capable of exploring well motivated the high mass parameter space, inaccessible to other experiments. With the formalism used here it is easy to check the power produced by such an experiment, which so far has only been calculated classically. For simplicity we will consider the medium to be isotropic and relatively large, such that boundary effects are unimportant.

As DM axion is nonrelativistic the velocity is negligible. Thus to first approximation it becomes impossible to distinguish the couplings to longitudinal or transverse photons. Because of this both contributions must be calculated in order to get a correct rate.

For DM axions with a distribution $f_a$ the absorption rate $\Gamma_{\rm abs}$ is simply given by
\begin{equation}
	\Gamma^{\rm axion}_{\rm abs}\simeq -f_a\frac{{\rm Im}\,{\Pi_{\rm axion}}}{\omega}\,.
\end{equation}
In writing this we assume that the occupation number of axions is large ($f_a\gg 1$), as well as being much larger than the occupation number of photons. This holds as a cryogenic plasma has few thermal photons and axion DM is highly occupied for $m_a\lesssim{\cal O}({\rm eV})$). In Appendix~\ref{TFTDM} we show that this thermal field theory approach is valid even if the axion dark matter is in a highly non-thermal state. The transverse part can be handled similarly to the longitudinal part studied above. Using the on-shell condition $K^2=m_a^2$ and treating the axion nonrelativistically, $\omega=m_a$, we see that
\begin{align}
	\Gamma^{\rm axion}_{\rm abs} =f_a\frac{ m_a^2 g_{a\gamma}^2B_{||}^2\Gamma_L}{(m_a^2-\omega_p^2)^2+(m_a \Gamma_L)^2}\nonumber\\
	+f_a\frac{ m_a^2 g_{a\gamma}^2B_{\perp}^2\Gamma_T}{(m_a^2-\omega_p^2)^2+(m_a \Gamma_T)^2} \, ,
\end{align}
where we have used equation \eqref{weldon} to define ${\Gamma_T=-{\rm Im}\,\Pi_T/\omega}$. As longitudinal and transverse photons are indistinguishable in the zero momentum limit,  ${\Gamma_T=\Gamma_L\equiv \Gamma_\gamma}$. As both contributions are now equal up to the projection of the magnetic field, we find on resonance ($m_a=\omega_p$) that
\begin{equation}
	\Gamma^{\rm axion}_{\rm abs}=f_a\frac{ g_{a\gamma}^2B^2}{\Gamma_\gamma}\,.
\end{equation}
As long as the axion-line width is much smaller than the line width of the resonance for the total abosrbed power axion DM can be treated as a delta function of $N$ axions,
\begin{equation}
    f_a=N(2\pi)^3\delta^3({\bf k})\,.
\end{equation}
Thus the power absorbed in a homogenous volume $V$ is simply given by
\begin{equation}
	P=V\int \frac{d^3{\bf k}}{(2\pi)^3}\omega  \Gamma^{\rm axion}_{\rm abs} = g_{a\gamma}^2B^2\frac{Q}{\omega}V\rho_a\,,\label{eq:plasmapower}
\end{equation}
	where ${Q=\omega/\Gamma_\gamma}$ is the ``quality factor" and $\rho_a$ is the local DM density. Equation \eqref{eq:plasmapower} is in exact agreement with Ref.~\cite{Lawson:2019brd} in the limit where boundary conditions are negligible (in their notation, the ``geometry factor" tends to unity). Thus we confirm the classical calculation of Ref.~\cite{Lawson:2019brd}, and see that for a sufficiently large medium so that boundary effects are unimportant both transverse and longitudinal polarizations play a significant role in the generated signal.
	\section{Conclusion}
 	In this paper we have reconsidered the calculation of axion emission from a thermal bath of photons using thermal field theory. As longitudinal plasmon-axion conversion is resonant, but largely neglected in the literature, many interesting physical environments are yet to be properly explored. In the interest of revitalising plasmon-axion conversion, we have applied our results to the most relevant astrophysical and laboratory targets, comparing the energy loss due to plasmon-axion conversion to other processes and to the present experimental bounds. In different energy regimes this new process dominates over Primakoff or bremsstrahlung processes.
	 
	We first considered the closest source of astrophysical axions, the Sun. While the luminosity of solar axions is largely set by the Primakoff effect, at low energies plasmon-axion conversion provides a new and dominant source of axions. This new flux motivates $1-100\,$eV-scale helioscope experiments.  Such an experiment would benefit from the improvements to optics at low energies. In the event of a discovery a low energy helioscope would provide an additional probe of stellar structure.
	 
	  For WDs this resonant conversion mechanism provides purely photonic contribution to the axion flux. For high mass WDs, the high inner density suppresses axion plasmon conversion, so only the outer shell contributes and the flux is subdominant relative to bremsstrahlung. It is possible that in low mass, high temperature and high magnetic field WDs plasmon production may dominate for electrophobic axions, leading to stronger bounds on the axion-photon coupling.  
	  
	  Lastly we used our thermal field theoretic calculations to confirm the behaviour of plasma haloscopes in the large medium limit, demonstrating that both transverse and longitudinal modes contribute to the absorption of dark matter axions.
	  
	   As plasmon-axion conversion is relatively unexplored compared to more traditional production mechanisms, there are still many astrophysical and laboratory environments to explore. A prime example is the magnetosphere of neutrons stars, where the density and the magnetic field strength are very promising. However the magnetic fields involved are too strong for the simple treatment outlined here and thus left for future work.
\\	   
	   	   
\section*{Acknowledgements}
% %%%%%%%%%%%%%%%%%%%%%%%%%%%%%%%%%%%%%%%%%%%%%%%%%%%%%%%%%%%%%%%%%%%%%%%%%%%%%
We are grateful to John Beacom, Chris Dessert, Luca Di Luzio, Mark Hollands, Kenny Ng, Ben Safdi and Aldo Serenelli for interesting discussions. We thank Leandro G. Althaus and Alejandro H. C\'orsico for providing the WD model. We thank Alessandro Mirizzi, Georg Raffelt and Javier Redondo for discussions and comments on the draft. 
AC acknowledges support from the “Generalitat Valenciana” (Spain) through
the “plan GenT” program (CIDEGENT/2018/019), as
well as national grants FPA2014-57816-P, FPA2017-
85985-P. AM is supported by the
European Research Council under Grant No. 742104 and is supported in part by the research environment grant ``Detecting Axion Dark Matter In The Sky And In The Lab (AxionDM)" funded by the Swedish Research Council (VR) under Dnr 2019-02337. The  work  of EV  was  supported  by the U.S. Department of Energy (DOE) Grant No.  DE-SC0009937. This research was supported by the Munich Institute for Astro and Particle Physics (MIAPP) which is funded by the Deutsche Forschungsgemeinschaft (DFG, German Research Foundation) under Germany's Excellence Strategy – EXC-2094 – 390783311.\\

\appendix
\section{Vertex factor for longitudinal polarization}
Here we derive the factors entering the vertices of the axion self energy, as used in Eq.~\eqref{selfenergy}. We restrict ourselves to the computation of the longitudinal component, which is the focus of the present work. An analogous calculation can be performed also for the transverse mode.

The Lagrangian is given by
\begin{equation}
{\mathcal L}=-  \frac{1}{2}  g_{a\gamma}a(\partial_{\mu}A_{\nu})\tilde{F}^{\mu\nu}\,. \label{lag}
\end{equation}
While it is tempting to use $A_{\nu}$ as the electric field, and $\tilde{F}^{\mu\nu}$ to provide the magnetic field, this would actually undercount how many ways one can assign these fields, leading to a reduction by a factor of two. To calculate the vertex factor, the easiest starting point is to rewrite Eq.~\eqref{lag} in terms of the electric and magnetic fields,
\begin{align}
    \frac{1}{2}(\partial_{\mu}A_{\nu})\tilde{F}^{\mu\nu}&=\frac{1}{4}\epsilon^{\mu\nu\rho\sigma}\partial_\mu A_\nu(\partial_\rho A_\sigma-\partial_\sigma A_\rho)\nonumber\\
    &=\frac{1}{2}\epsilon^{ijk}\partial_0A_i\partial_jA_k-\frac{1}{2}\epsilon^{ijk}\partial_iA_0\partial_jA_k \nonumber\\
    &+\frac{1}{2}\epsilon^{ijk}(\partial_0A_k-\partial_kA_0)\nonumber\\
    &=-{\bf E}\cdot {\bf B}\,,
\end{align}
where we have used that $E_i=F_{0i}$ and $B_i=\frac{1}{2}\epsilon_{ijk}F^{jk}$.
This writing allows us to unambiguously assign ${\bf B}$ to be the external magnetic field. We can now calculate the contribution to the vertex factors,
\begin{align}
    g_{a\gamma}{\bf E}\cdot{\bf B}&=-g_{a\gamma}\left(\frac{d{\bf A}}{dt}+{\bf \nabla}A^0\right)\cdot {\bf B}\nonumber\\
    &=g_{a\gamma}\left(\frac{\omega^2{\bf \hat k}}{\sqrt{\omega^2-k^2}}-\frac{k^2{\bf \hat k}}{\sqrt{\omega^2-k^2}}\right)\cdot {\bf B}\,e^{{\bf k}\cdot{\bf x}-\omega t}\nonumber\\
    &=g_{a\gamma}m_a{\bf \hat k}\cdot{\bf B}\,e^{{\bf k}\cdot{\bf x}-\omega t}\,,
\end{align}
where in the second line we used that the longitudinal plasmon is a plane wave with polarization vector given by Eq.~\eqref{Polarisation} and, as we are considering the axion self energy, $\sqrt{\omega^2-k^2}=m_a$. Note that there are two vertices entering Eq.~\eqref{selfenergy}, leading to a mod squaring.

\section{Axion to photon conversion}
\label{Appendix: conversion}
Here we report for completeness the treatment for axion to photon conversion in the external magnetic field of the WD~\cite{Dessert:2019sgw,Raffelt:1987im}. The probability can be found working in the small mixing approximation and using time independent perturbation theory. The axion-photon conversion probability is~\cite{Hook:2018iia} 
 \begin{equation}
 	p_{a\rightarrow \gamma}=\left| \int^{\infty}_{R_{\rm{WD}}}dr' \Delta_{B}(r')e^{i \Delta_a r' -i \int^{r'}_{R_{\rm{WD}}}dr'' \Delta_{\parallel}(r'')}\right|^2\,,
 \end{equation}
 where one integrates from the surface of the WD to infinity.  In the above expression we used the terms
   \begin{subequations}
\begin{alignat}{2}
 \Delta_B(r)&=(g_{a\gamma}/2B(r))\sin(\Theta)\,,\\
  	\Delta_{||}(r)&=7/2 E   (\alpha_{em}/45\pi)(B(r)/B_{\rm crit})^2 \sin\Theta^2\,,
\end{alignat}
\end{subequations}
and
 \begin{equation}
  	\Delta_a= -m_a^2/2\omega\,,
  \end{equation}
  with ${B_{\rm crit}\simeq 4.41\times 10^{13}}\,$G and $\Theta$ the angle between the radial propagation direction and the magnetic field. \\
  
\section{Primakoff process}\label{appendix: primakoff}

The Primakoff process which occurs in stars is the conversion $\gamma \leftrightarrow a$ in the presence of electric fields of nuclei and electrons. The Primakoff process is most relevant in the Sun, where the conditions are non-relativistic so that both electrons and nuclei can be treated as heavy with respect to the scattering photon. Considering a target with charge $Z e$, the differential rate reads~\cite{Raffelt:1985nk, Jaeckel:2006xm}
\begin{equation}
    \frac{d\Gamma_{\gamma \rightarrow a}}{d\Omega} = \frac{g_{a\gamma}^2Z^2\alpha}{8\pi}\frac{|\textbf{k}_a \times \textbf{k}_{\gamma}|^2}{\textbf{q}^4} \frac{\textbf{q}^2}{\textbf{q}^2 + k_{\rm S}^2}\,,
\end{equation}
where $\textbf{k}_{\gamma}, \, \textbf{k}_{a}$ are the spatial momenta of the photon and the axion, while $\textbf{q} = \textbf{k}_{\gamma} - \textbf{k}_{a}$ is the momentum transfer. The factor $\frac{\textbf{q}^2}{\textbf{q}^2 + k_{\rm S}^2}$ comes from Debye screening~\cite{Raffelt:1985nk}. We neglect temporal variations in the electric field of the heavy particle, meaning that the axion and photon have the same energy~\cite{Raffelt:1987np}. 

Defining $\theta$ as the angle between $\textbf{k}_{\gamma}$ and $\textbf{k}_{a}$ we can then write~\cite{Jaeckel:2006xm}
\begin{align}
    \Gamma_{\gamma \rightarrow a}&= \frac{g_{a\gamma}^2 k_{\rm S}^2T}{16\pi}\int^{1}_{-1}d\cos\theta \frac{k_{\gamma}^2k_{a}^2\sin^2\theta}{k_{\gamma}^2 + k_{a}^2 - 2 k_{\gamma}k_a \cos\theta} \nonumber \\ &\times \frac{1}{k_{\gamma}^2 + k_{a}^2 - 2 k_{\gamma}k_a \cos\theta + k_{\rm S}^2}\nonumber\\
    &=\frac{g_{a\gamma}^2k_{\rm S}^2T}{64\pi}\int^{1}_{-1}d\cos\theta\frac{\sin^2\theta}{(x-\cos\theta)(y-\cos\theta)}\,,
\end{align}
where $x=(k_a^2+k_\gamma^2)/2k_ak_\gamma$ and $y=x+k_{\rm S}^2/2k_ak_\gamma$. The factor of $k_{\rm S}^2T$ comes from summing over all possible particle species and rewriting the weighted factors of $Z^2\alpha$. The energy-loss rate per unit volume then reads 
\begin{equation}
    Q = \int \frac{2 d^3\textbf{k}_{\gamma}}{(2\pi)^3} \frac{\omega \, \Gamma_{\gamma \rightarrow a}}{\exp(\omega/T(r))-1}\,.
\end{equation}
Notice that the rate $\Gamma_{\gamma \rightarrow a}$ is a function of the energy and the position in the star, as the Debye screening length and the plasma frequency vary from point to point. As usual one can then get the total luminosity integrating over the volume. As we are primarily concerned with ultrarelativistic axions, we neglect the axion mass giving $k_a \simeq \omega$. The dispersion relation for transverse plasmons $k_{\gamma}^2(r) = \omega^2 -\omega_{p}^2(r)$ imposes the restriction that only axions with $\omega\geq\omega_p(r)$ are produced. 

\section{Thermal field theory and axion dark matter}\label{TFTDM}
We wish to apply the formalism of thermal field theory to DM axions being absorbed by a cyrogenic plasma. However, axion DM most likely does not exist as a thermal state, and can in principle be far from equilibrium. Here we show that our formalism can indeed be consistently applied to such a case. For a bosonic particle species with distribution $f$, absorption rate $\Gamma_{\rm abs}$ and production rate $\Gamma_{\rm prod}$ one can show~\cite{Weldon:1983jn}
\begin{equation}
    \frac{\partial f}{\partial t}=-f\Gamma_{\rm abs}+(1+f)\Gamma_{\rm prod}\,.
\end{equation}
For our purposes, we are interested in the coupled axion-photon system, which to good approximation only has single particle production. For axion DM with $m_a\lesssim{\cal O}({\rm eV})$ the occupation number of axions is very high, giving $f_a\gg 1$. We then see that
\begin{subequations}
\begin{align}
    \frac{\partial f_a}{\partial t}&\simeq-f_a\Gamma^{\rm axion}\\
    \frac{\partial f_\gamma}{\partial t}&=-f_\gamma\Gamma^{\rm axion}_{\rm prod}+(1+f_\gamma)\Gamma^{\rm axion}_{\rm abs}\,,
\end{align}
\end{subequations}
where $\Gamma^{\rm axion}=\Gamma^{\rm axion}_{\rm abs}-\Gamma^{\rm axion}_{\rm prod}$.
As we are only considering conversions between axions and photons, $\partial f_\gamma/\partial t=-\partial f_a/\partial t$ giving
\begin{equation}
    \Gamma^{\rm axion}_{\rm abs}\simeq(f_a-f_\gamma)\Gamma^{\rm axion}\simeq -f_a\frac{{\rm Im}\Pi_{\rm axion}}{\omega}\,,
\end{equation}
where in the last approximation we have assumed that $f_a\gg f_\gamma$ and used Eq.~\eqref{weldon}. Note that for axion-photon conversion in a magnetic field one does not need to worry about Bose stimulation factors, in general any enhancement from Bose stimulation is canceled by the same enhancement in back-reaction, giving exactly the result shown here~\cite{Ioannisian:2017srr}. Thus for a sufficiently highly occupied axion DM state we can apply our thermal field formalism used throughout the paper. 
\bibliographystyle{bibi}
\bibliography{biblio.bib}
\end{document}